\documentclass[prl,aps,twocolumn,superscriptaddress,showpacs]{revtex4}
\usepackage{amsfonts}
\usepackage{psfrag}
\usepackage{graphicx}
\usepackage{graphics}
\usepackage{bm}
\usepackage{color}
\usepackage{epsf,epsfig,pdfpages}
\usepackage{verbatim,color,ulem}
\newcommand{\be}{\begin{equation}}
\newcommand{\ee}{\end{equation}}
\newcommand{\ba}{\begin{eqnarray}}
\newcommand{\ea}{\end{eqnarray}}

\newcommand{\dcom}[1]{}
\newcommand{\dnote}[1]{}

\newcommand{\gsim}{\raise.3ex\hbox{$>$\kern-.75em\lower1ex\hbox{$\sim$}}}
\newcommand{\lsim}{\raise.3ex\hbox{$<$\kern-.75em\lower1ex\hbox{$\sim$}}}

\begin{document}

\title {Quantum sound-cone fluctuations in cold Fermi gases: Phonon propagation}
\author{Jen-Tsung Hsiang}
\affiliation{Department of Physics,
National Dong-Hwa University, Hualien, Taiwan, R.O.C.}
\affiliation{Center for Theoretical Physics, Fudan University,
Shanghai, China}
\author{Da-Shin Lee}
\affiliation{Department of Physics,
National Dong-Hwa University, Hualien, Taiwan, R.O.C.}
\author{Chi-Yong Lin}
\affiliation{Department of Physics,
National Dong-Hwa University, Hualien, Taiwan, R.O.C.}
 \author{Ray\ J.\ Rivers}
\affiliation{ Blackett Laboratory, Imperial College
London, SW7 2BZ, U.K.}

\date{\today}
\begin{abstract}
We examine the effect of quantum fluctuations in a tunable cold
Fermi gas on phonon propagation. We show that these fluctuations can
be interpreted as inducing a stochastic space-time. This
effect can be displayed in the variation in the travel time of
phonons, at its greatest in the crossover region between
BEC and BCS regimes.
\end{abstract}


\maketitle
Massless particles and gapless modes (photons, phonons) propagate causally according to the metrics of their respective light- and sound-cones. However, the intrinsically quantum mechanical nature of their environments (quantum gravity, condensate fluctuations) makes the cones 'fuzzy'. In particular, this induces fluctuations in their times of flight. Such an effect is difficult to calculate {\it ab initio}.  Instead, several authors in the last few years
have adopted the simpler position of deriving this and related effects  for photons \cite{hu,ford} and phonons \cite{gurarie,krein,gaul} in phenomenological random media.
In this paper we show that, for very cold tunable Fermi gases, the calculable fluctuations in the diatom density provide the random medium with which the phonons scatter. From these the quantum fluctuations in phonon times of flight can be determined. Roughly, they are somewhat less than one percent effects on the propagation time of waves across a typical condensate. Nonetheless, this is huge in comparison to the relative $10^{-9}$ fluctuations in photon propagation times in random media \cite{ford}, which are their nearest equivalent, let alone the infinitesimally small Planck time induced by the fluctuations of quantum gravity, which prompted the analysis.

We adopt the notation of our earlier work \cite{lee1,lee2} in
describing a cold ($T=0$) Fermi gas, tunable through a narrow
Feshbach resonance, by the action ($\hbar = 1$) \cite{gurarie2}
\begin{eqnarray}
S &=& \int dt\,d^3x\bigg\{\sum_{\uparrow , \downarrow}
 \psi^*_{\sigma} (x)\ \left[ i \
\partial_t + \frac{\nabla^2}{2m} + \mu \right] \ \psi_{\sigma} (x)
\nonumber \\
   &+& \varphi^{*}(x) \ \left[ i  \ \partial_t + \frac{\nabla^2}{2M} + 2 \mu -
\nu \right] \ \varphi(x) \nonumber \\
&-& g \left[ \varphi^{*}(x) \ \psi_{\downarrow} (x) \ \psi_{\uparrow}
(x) + \varphi(x)  \psi^{*}_{\uparrow} (x) \ \psi^{*}_{\downarrow} (x)
\right]\bigg\} \label{Lin}
\end{eqnarray}
 for fermion fields $\psi_{\sigma}$
 with spin label $\sigma = (\uparrow, \downarrow)$. The diatomic field $\varphi$ describes the bound-state (Feshbach) resonance with
 tunable binding energy $\nu$ and mass $M =2m$, and
 $ - g\,\varphi (x) = g|\varphi(x)| \ e^{i\theta (x)}$
 represents the  condensate.

As a result of spontaneous symmetry breaking a homogeneous condensate acquires a non-zero
 $|\varphi (x)|=|\varphi_0|$.
We expand  in the derivatives of $\theta$ and the {\it small}
fluctuations in the condensate density
 $\delta|\varphi| = |\varphi|  -  |\varphi_{0}|$,
 always preserving the Galilean invariance of the system. Galilean scalars are the
density fluctuation
 $\delta |\varphi|$ itself,
$G(\theta) = \dot{\theta} + (\nabla
\theta )^2/4m$, and  $D_t(\delta |\varphi|,\theta ) =
\dot{(\delta |\varphi|)}+ \nabla \theta .\nabla (\delta |\varphi|)/2m$, the comoving time derivative in the condensate with fluid velocity $\nabla \theta/2m$.

The action ${S}$ is quadratic in the fermion fields. On integrating them out and changing variables to $\theta$ and $\epsilon = \kappa^{-1}\delta|\phi|$,
a {\it dimensionless} rescaled condensate fluctuation, the {\it local} Galilean invariant effective
density for the long-wavelength, low-frequency condensate is of the form \cite{lee1,lee2}
\begin{eqnarray}  \label{LeffU0}
 S_{\rm eff}[\theta, \epsilon] &=&
 S_{0}[\theta ]
-\alpha\int d^4x~\epsilon G(\theta)
\\
&+& \frac{1}{4}\int d^4x [\eta {\dot\epsilon}^{ 2}
-\rho_0(\nabla\epsilon)^2/2m - {\bar M}^2\epsilon^{ 2}],
\nonumber
 \end{eqnarray}
 where
 \vskip -0.6cm
 \be
 S_{0}[\theta ] =\!\! \int d^4x\bigg[\frac{N_0}{4}\ G^2(\theta) -\frac{1}{2}{\rho}_0
   G(\theta)   \bigg]
 \label{S0}
 \ee
is the canonical acoustic
BCS action \cite{aitchison}.
\noindent{
 The scale factor $\kappa$
is chosen so that the coefficients of $(\nabla\epsilon)^2$ and $(\nabla\theta)^2$ in (\ref{LeffU0}) and (\ref{S0}) are identical \cite{aitchison}. The coefficients $\alpha, \eta$, etc. are known functions of the scattering length \cite{lee1,lee2} and hence of the external magnetic field used to tune the condensate from the BCS to BEC regimes.
The action (\ref{LeffU0}) represents a two-component
system of molecules and atom pairs, with a corresponding
    two-component density in which fermions  oscillate from one to the other while maintaining a fixed total number density $\rho_0$.
In the hydrodynamic approximation, where the spatial and temporal
variation  of $\epsilon$ can be ignored in comparison to $\epsilon$
itself, density fluctuations $\epsilon$ act as sources and sinks to
the dynamics of the phase $\theta$ and can be eliminated by simply
identifying ${\epsilon}\approx
 - 2{\alpha} G(\theta)/{\bar M}^2$.  The corresponding
  Euler-Lagrange equation for $\theta$
 is the continuity equation of a {\it
single} fluid \cite{lee1,lee2}
  from which the fluctuations in the local number density $\delta \rho  = \rho - \rho_0 = 2\alpha \epsilon -N_0 G(\theta )= - (N_0+ 4 \alpha^2/ \bar M^2) G(\theta )$ and the number current density ${\bf j} = \rho_0 \nabla\theta/2m$ lead to the wave
equation:
\begin{equation}
\ddot \theta (x) -c^2  \nabla^2\theta (x)=0 \, .
\end{equation}
%
%That is, the sound
%wave can be created by classical perturbations of  the number
%density in the condensates.
The sound speed $c$ will be derived below.

We see immediately that beyond the
hydrodynamical approximation $\epsilon$ becomes a
dynamical field with quantum fluctuations to which the phonons
couple.  Coarse-graining the $\epsilon $ field will introduce
stochasticity in the acoustic metric of the $\theta$ field via its
Langevin equation.

We proceed by constructing the closed
time-path (CTP) effective action,
\ba
&& \!\!\!\!\! S_{\rm
CTP}[\theta^+, \epsilon^+ ; \theta^-, \epsilon^-] = S_{\rm eff}[\theta^+, \epsilon^+] - S_{\rm eff}[\theta^-, \epsilon^-]
\nonumber
\label{SNONEQeff}
\ea
\vskip -0.1cm
\noindent
{where $\pm$ denote integration on the upper
and lower contours of the path respectively.} It is sufficient to retain only the second power of
$\epsilon$. Integrating out
the $\epsilon$ field (e.g. see \cite{boyanovsky}) then gives an
effective non-local action for dynamical phonons,
\ba
S_{\rm eff}[\theta^+; \theta^-] =
S_{0}[\theta^+ ] - S_{0}[\theta^- ] + \Delta S[\theta^+; \theta^-]
\label{CTP}
\ea
where 
\ba
&& \hspace{0.2cm} \Delta S[\theta^+; \theta^-] =
\\
\label{CTP}
%\ba &&
&&\hspace{-0.4cm} \frac{\alpha^2}{2}\int \int~d^4x_1 d^4x_2  
\!\!\!\!\!
\sum_{a, b= +, -}
\!\!\! G(\theta^a(x_1))D^{ab}_{\epsilon}(x_1 - x_2)G(\theta^b(x_2)) 
\nonumber
\ea
In (\ref{CTP}) the
$D^{\pm\pm}_{\epsilon}(x_1 - x_2)$
denote the $\epsilon$ correlators
\ba
&&D^{++}_{\epsilon} = \theta(t_1-t_2) \, \langle \epsilon (x_1)
\epsilon (x_2) \rangle + \theta(t_2-t_1) \, \langle \epsilon (x_2)
\epsilon (x_1) \rangle  \nonumber\\
&&D^{--}_{\epsilon} = \theta(t_1-t_2) \, \langle \epsilon (x_2) \epsilon
(x_1) \rangle + \theta(t_2-t_1) \, \langle \epsilon (x_1)
\epsilon (x_2) \rangle \nonumber\\
&&D^{+-}_{\epsilon} = - \langle \epsilon (x_2) \epsilon (x_1) \rangle =
D^{-+}_{21} .\label{noneq_greenfun}
\ea

We recover the semiclassical phonon field $\theta$ and the
fluctuating field $R$ about it through the decomposition
$\theta^{\pm}(x) = \theta (x) \pm R/2. $ For the purpose of wave
propagation we need only to retain terms in $S_{eff}$
linear and quadratic in $R$. Quadratic terms in $R$
are then linearised by the introduction of noise $\xi$ enabling us
to extend
$S_{eff}[\theta^+; \theta^-] \equiv S_{eff}[\theta; R]$
to the form
\be
S_{\rm eff}[\theta; R,\xi] = \int d^4x~R(x)~L(\theta,
\partial_{\mu}\theta, \xi)(x).
\ee
$R$ is now understood as a Lagrange multiplier
to the Langevin equation $ L(\theta,
\partial_{\mu}\theta, \xi) = 0 $ describing the propagation of
phonons in a stochastic background provided by the noise.

Specifically, $G(\theta^{\pm})\approx
\dot{\theta}~{\pm}~D_tR/2$
at the relevant order. $\Delta S$ in (\ref{CTP}) splits into
two terms, linear and quadratic in $R$ respectively. The linear term
will give the modification on the dynamics of the phonons obtained
from retardation effects through the $\epsilon$ retarded propagator
$D_{\epsilon R}$,
\be
 D_{\epsilon R} (x-x') = i \theta(t-t') \langle [ \epsilon(x), \epsilon(x') ]
 \rangle \, .
 \label{D_R}
 \ee
On introducing  a bilinear coupling $\alpha
(D_t R(x))\xi(x)$ the quadratic term  can be rewritten in terms of  the Gaussian
noise $\xi$ with distribution
\ba
\langle\xi(x)\xi(x')\rangle &=& D_{\epsilon H}(x-x')=
\frac{1}{2} \langle \{ \epsilon(x), \epsilon(x') \}
 \rangle. \nonumber\\
&=&  \int \frac{d^3{\bf k}}{( 2\pi)^3}
 \frac{\cos[{\omega}_k (t-t')]}{{\omega}_k {\eta}} \, e^{-i {\bf k} \cdot ({ \bf x}-{\bf x'})}
 \,
 \label{D_H}
\ea
in which the dispersion relation of the $\epsilon$ field is
determined by
$\omega_k =\sqrt{\rho_0 k^2/2m \eta +\bar M^2/\eta}$.

On integrating by parts
 $(D_t R(x))\xi (x)\rightarrow - R(x)( \dot\xi + \nabla . (\xi \nabla \theta)/ 2m)(x)=
 - R(x)( D_t \xi + \xi \nabla^2 \theta)/ 2m)(x)$,
the resulting Langevin equation is
then
\ba
&&\!\!\!\!\!\!\!\!\!\!\!\!
\frac{N_0 }{2}{\ddot\theta}(x)
\!-\!(\frac{\rho_0}{4m} \!-\! \frac{\alpha\xi}{2m})
   (\nabla^2\theta)
 \!+\!\alpha^2\! \int\! d^4x'
 \partial_t D_{\epsilon R} (x-x') \, {\dot\theta}(x')
 \nonumber
 \\
  && =  -\alpha {D_t \xi}(x).
%\nonumber
\label{lang4}
\ea
\vskip -0.1cm
\noindent
What is crucial for our subsequent discussion is the multiplicative
noise term $\xi \nabla^2 \theta$, a consequence of the Galilean
invariance enforcing covariant derivatives.
Behaviour of this form  is the starting point for
the stochastic analysis of the papers of
\cite{hu,ford,gurarie,krein}. However, whereas these authors argue
for stochastic behaviour on empirical grounds, in our case  we see
from (\ref{D_H}) that the noise $\xi$ is essentially the (known)
fluctuation field $\epsilon$.

Eq.(\ref{lang4}) encodes quantum effects in two distinct ways, through the retarded commutator $D_{\epsilon R}$ and the noise $\xi$. Although they overlap we shall do our best to treat them separately.

Firstly, for comparative purposes, let us neglect $\xi$ in
(\ref{lang4}). In the  phonon acoustic limit $\omega = ck$ for
which, as $\omega, k\rightarrow 0$ in (\ref{D_R}), $ D_{\epsilon R}
(x - x')\rightarrow (2/{\bar M}^2)\delta^4(x -x'),$  we reproduce
the classical mean value speed of sound $c$:
\be
c^2 = \frac{\rho_0/2m}{N_0 + 4\alpha^2/{\bar M}^2}.
\label{c2}
\ee
\vskip -0.1cm
If $a_S$ is
the s-wave scattering length and $k_F$ the Fermi momentum, $c^2/v_F^2$ varies smoothly with $1/k_F a_S$, decreasing monotonically from 1/3 in the  BCS regime ($1/k_F a_S < 0$) to vanishingly small in the BEC regime ($1/k_F a_S < 0$) \cite{lee1,lee2}.
%\begin{figure}
%\centering
%\includegraphics[width=0.6\columnwidth=0.8]{2015_01_21_Fig1.pdf}
%\includegraphics[width=7cm]{Fig2.ps}
%\caption{\textcolor{blue}{The Figure shows the behavior of
%$c^2/v_F^2$ as a function of $1/a_S k_F$ for $\gamma_0\approx 0.6$
%and ${\bar g}^2$ (see text). The inset shows the variation of the
%momentum scale $K$ of Eq.(\ref{K}) also as a function of $1/a_S
%k_F$.}}
%\end{figure}
%
 More generally, if we take $ D_{\epsilon R} (x)$ as follows from (\ref{D_R})  we have a Bogoliubov quantum 'rainbow' of sound speeds
$c_k$, according to the wavelength $k$, of the form \cite{lee1,lee2}
 \begin{eqnarray}
  c_k^2 &\approx&  c^2[ 1 + k^2/K^2 + ...] \, ,
  \label{cklin}
 \end{eqnarray}
where
\vspace{-0.3cm}
 \begin{equation}
 K^{-2} = \frac{4\alpha^2 c^2}{{\bar M}^4}\left[ 1- \frac{c^2 \eta}{\rho_0/2m}\right] \, .
 \label{K}
 \end{equation}
\vskip -0.0cm
In the large momentum limit in the BEC regime we recover \cite{lee2} the free particle limit for diatoms/molecules
$\omega = k^2/4m = k^2/2M$. Provided that the phonons comprise a wavepacket propagating
 toward the detector with  central momentum $k_0$ and width $\Delta k_0$, with $k_0 + \Delta k_0 < K$ of (\ref{K})
they all experience approximately the same sound speed $c$ and our
estimates for fluctuations in times of the flight are unchanged.
This we now assume (see inset (top) in Fig.1).

For such long-wavelength phonons equation (\ref{lang4})
becomes
 \be
 {\ddot\theta}(x) - c^2(1 - 2\alpha\xi/\rho_0)\nabla^2\theta\approx  - 4m(\alpha/\rho_0) c^2 {D_t \xi}(x).
 \label{langac2}
 \ee
in terms of the speed of sound $c$ of (\ref{c2}).
As a result we can interpret $c_{\xi}$,
\be
 c_{\xi}^2 = c^2 (1 - 2\alpha\xi/\rho_0),
 \label{cstoch}
\ee
 as a stochastic speed of sound in the long wavelength regime.

Our main interest is the effect of the fluctuating background on the propagation of photons. We follow
the analysis of \cite{hu,ford}. For a spatially homogeneous static condensate its operator-valued
acoustic metric can be taken as $ dt^2 - c_{\xi}^{-2} {\bf dx}^2 =
0, \label{d2} $
 written
as
\be
c^2 dt^2 - (1+ 2\alpha\xi/\rho_0){\bf dx}^2 = 0.
\ee
 In conventional formalism the phonon propagates along the sound cone determined by the
null-geodesic:
\begin{equation}
c^2 d t^2= d {\bf{x}}^2 +h_{ij} dx^i dx^j \, ,
\end{equation}
where $h_{ij} = (2\alpha/\rho_0)\xi\delta_{ij}$.
If the spatial
separation between the source and the detector is $r$, then the
travel time can be expressed  as
\begin{equation}
T = \int_0^T dt \approx \int_0^r dr \frac{1}{c} \bigg[ 1+ \frac{1}{2}
h_{ij} n^i n^j \bigg] \, ,
\end{equation}
where  $ dr= d|{\bf x}|$ and ${\bf{n}}^i= dx^i /dr$ is a unit vector
along the direction of the sound wave propagation. The local
velocity $c$ is evaluated on the unperturbed path of the waves
$r(t)$, which we take along the $z$- direction, so that $z(t)=c t$. With $\langle h_{ij}\rangle = 0$, the variance of the
travel time is given by
\begin{eqnarray}
&&
(\Delta T)^2 = \langle T^2 \rangle-\langle T\rangle^2
\nonumber
\\
&&\!\!\!\!\!\!
=\frac{1}{4} \int_0^{ T }\!\! d t_1 \!\! \int_0^{ T }\!\! d t_2 \,
 n^i n^j n^l n^m \langle h_{ij} (r(t_1),t_1) \,
h_{lm}(r(t_2),t_2) \rangle \, \nonumber \\
&=& \!\! \frac{\alpha^2}{\rho_0^2}\int_0^{ T } d
t_1 \int_0^{ T } d t_2 \langle \xi (z(t_1),t_1) \, \xi(z(t_2),t_2)
\rangle. \,
\label{T}
\end{eqnarray}
With the noise correlation given by $D_{\epsilon H}$ in (\ref{D_H}),
Eq.(\ref{T}) is our key result, but to see whether it can be tested
is not straightforward. We have in mind an experiment along the
lines of that described in \cite{andrews}, discussed further in
\cite{kavoulakis}, in which sound pulses are created by density
perturbations. [Note that, for our condensate $c^2\propto \rho_0$
just as for a condensate of elementary bosons.]

Straightforward substitution of $D_{\epsilon H}$ in
(\ref{D_H}) gives
\ba
&&
\hspace{-0.5cm}
 \left(\Delta T \right)^2
= \frac{\alpha^2}{\rho_0^2} \int_0^T\!
dt_1\!\int_0^T \! dt_2 \nonumber\\
&&
\hspace{-0.5cm}
\quad  \times \int^{k_{\Lambda}}_{0} \frac{k^2 d k}{4\pi^2}
 \frac{\cos[{\omega}_k (t_1-t_2)]}{{\omega}_k {\eta}} \frac{2 \sin[ck \vert t_1-t_2 \vert]}{k c \vert t_1-t_2 \vert }.
\label{deltat/t_exact}
\ea
$\left(\Delta T \right)^2$ shows a
logarithmic UV divergence because of the acoustic approximation and
we cut off momentum at $k = k_{\Lambda} = O(K) = O(k_F)$ in the relevant regime. The initial growth
of $\left(\Delta T \right)^2$ from zero at time zero is rapid, and
when $T_s \sim 1/\omega_{(k=0)}
=\sqrt{\eta}/\bar{M}$, the growth halts and $\left(\Delta T \right)^2$  saturates to its late time value.

For large times the $k$-integral in
(\ref{deltat/t_exact}) is dominated by large $k$ contributions and
can be approximated well by
\ba
&&
\hspace{-0.5cm}
 \left(\epsilon_F\Delta T\right)^2 \approx
\frac{\alpha^2}{\rho_0^2} \frac{(mv_F^2)^2}{4\pi^2 \eta \,
c^3}\frac{x^2}{(1-x^2)}
\nonumber
\\
&& 
\hspace{-0.5cm}
\big[ -\tanh^{-1} [ {x}/{\sqrt{1+y^2}} ] +
x\ln[{1+\sqrt{1+y^2}}/{y}] \; \big] \,, \label{deltat/t6}
\ea
where $y=\sqrt{2 m \bar M^2/\rho_0 k_\Lambda^2}$ and
$x=c/\sqrt{\rho_0/2m\eta}<1$ across the whole regime from
BCS to BEC. We stress that the behaviour described above is a
consequence of quantum fluctuations and not thermal fluctuations.

Before our numerical study,  we need to list the basic attributes of
the parameters in the model. [See \cite{lee1,lee2} for more detail.]
 We find that $0\leq
\alpha/\rho_0\leq 1$ increases as we tune the gas from the deep BCS
regime ($1/k_F a_S < 0$), when $\alpha/\rho_0\approx 0$ to the deep
BEC regime ($1/k_F a_S > 0$), when $\alpha/\rho_0\approx 1$. On the
contrary, $\eta, N_0$ and ${\bar M}^2$ go from finite values to zero
as we go from deep BCS to BEC regimes, in each of which $\eta\approx N_0$. As a result of ${\bar M}^2$
vanishing $c^2$ falls off from $v_F^2/3$ in the BCS regime
\cite{aitchison} to {\it zero} in the deep BEC regime
\cite{lee1,lee2}.  It follows that, for Fermi energy $\epsilon_F$, $\left(\epsilon_F\Delta T\right)^2\rightarrow 0$ in the deep BEC regime.  Also,
$\left(\epsilon_F\Delta T\right)^2\rightarrow 0$ in the deep BCS
regime since $\alpha \approx 0$ there

To be concrete, consider a cold $^6Li$ condensate of $ 3\times 10^5
$ atoms tuned by the narrow resonance at $H_0 = 543.25 G$, discussed
in some detail in \cite{strecker} and used by us elsewhere
\cite{lee1,lee2}. The narrowness of the resonance is best determined
by the dimensionless width
 $\gamma_0\approx\sqrt{\Gamma_0/\epsilon_F}$, where the resonance
$\Gamma_0$   \cite{gurarie2} is mainly given by $H_{\omega}$,
the so-called "resonance width"  of the central field $H_0$ required
to achieve infinite scattering length (the unitary limit).

We take the number density $\rho_0
=k_F^3/3 \pi^2\approx 1 \times 10^{11} cm^{-3}$~\cite{strecker}, for
which $\epsilon_F\approx 7 \times 10^{-12} eV $ ($\epsilon_F/\hbar
\approx 10~ ms^{-1}$) and $\gamma_0\approx 0.6$. In terms of the
dimensionless coupling $\bar g$, where $g^2 = (64\epsilon^2_F/3
k_F^3){\bar g}^2$ \cite{gurarie2}, $^6Li$ at the density above
corresponds to ${\bar g}^2 = 0.8$.
In the inset (bottom) to Fig.1 we plot the saturation value of
$\left(\epsilon_F\Delta T\right)^2$ obtained
from~(\ref{deltat/t_exact}) on varying $1/k_F a_S$, where we take the
UV cutoff $k_{\Lambda}= k_F$ for carrying out the
momentum integration in~(\ref{deltat/t_exact}) numerically.
 The maximum travel time fluctuation occurs near the crossover regime at $1/a_S k_F \approx
0.7$. The main figure in Fig.1 shows the evolution of $\Delta T$ for
this value of $1/a_S k_F$, achieving its saturation value of $\Delta
T\approx 0.9 \epsilon_F^{-1} \approx 0.1 ms$, in agreement with
Eq.(\ref{deltat/t6}), after $T_s \sim \sqrt{\eta}/\bar{M} \approx
1.0 \epsilon_F^{-1} \approx 0.1 ms$.  In particular, (see upper
inset) $K \approx 0.3 k_F$ at $1/a_S k_F \approx 0.7$ with the
central momentum $k_0 \approx 0.1 k_F$ determined by the sound speed
$c \approx 0.1 v_F$. With $k_F
\approx 1 /\mu m $ the width of the density fluctuations can be of
order several $\mu m$ moving on a condensate of size
$L \approx 100 k_F^{-1} \approx 100 \mu m$. With
$c\approx 1.4 \, \mu m/ms$ the time of flight from the centre is
approximately $30 ms$,} whence the one percent or less fluctuation effect
cited initially. Unfortunately, the effect is not yet testable since
experimentalists most easily measure the (saturated) fluctuations
$\Delta r = c\Delta T\approx 0.14\mu m$  in the position of the
propagating wavefront. Currently, such uncertainty is well within the
noise  by between one and two orders of
magnitude \cite{joseph}.

\begin{figure}
\centering
\includegraphics[width=0.96\columnwidth=0.8]{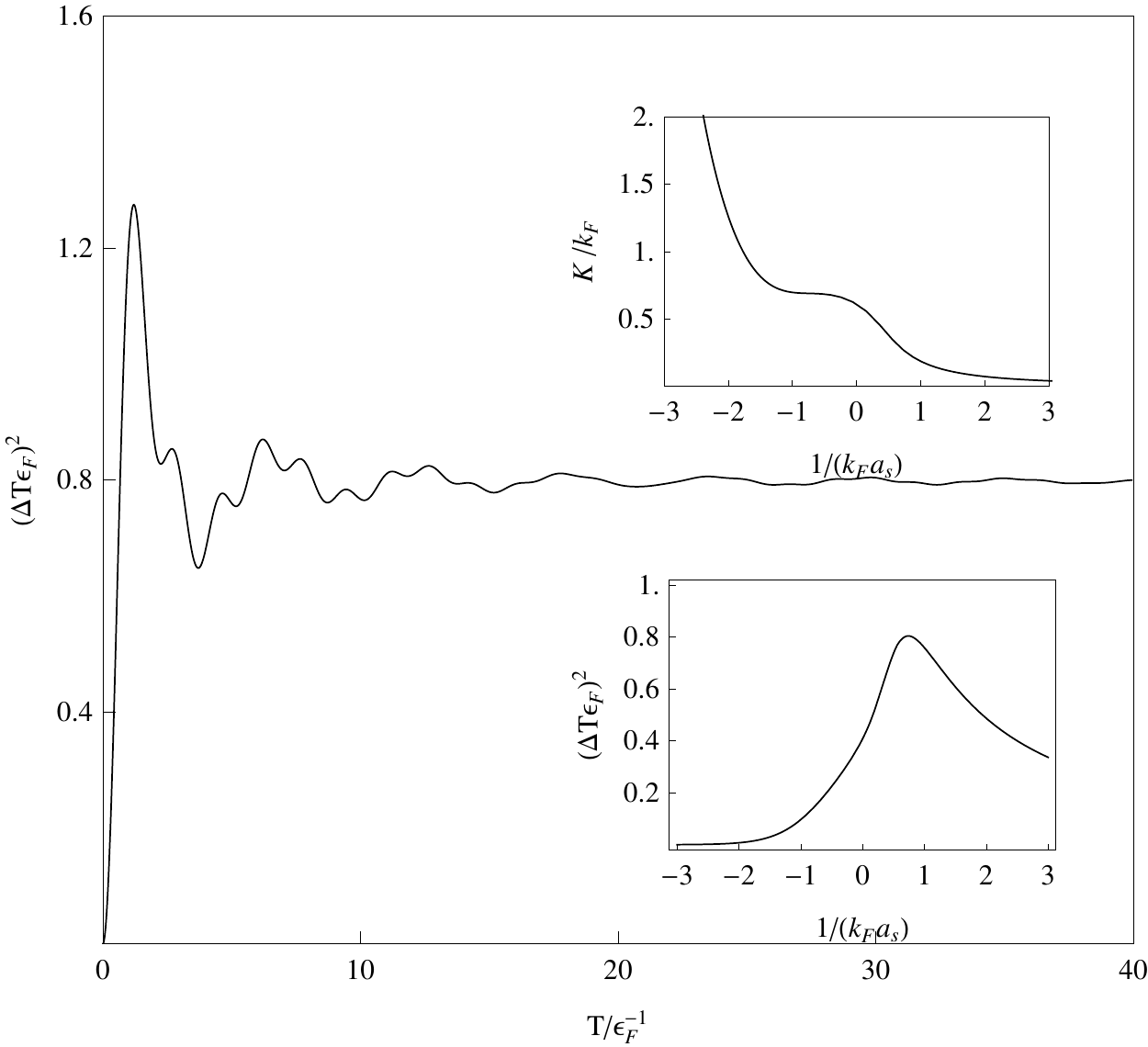}
\caption{The Figure shows the behavior of $\left(\epsilon_F \Delta
T\right)^2$ as a function of  $\epsilon_F T$  at $1/a_S k_F=0.7$,
given by (\ref{deltat/t_exact}), with the UV cutoff  $k_{\Lambda}=
k_F$ for numerically carrying out the momentum integration.  The lower
inset Figure shows the saturation value of fluctuations in
time of the flight by changing $1/a_S k_F$, also obtained from
(\ref{deltat/t_exact}).  Its maximum value occurs at $1/a_S k_F
=0.7$ near the crossover regime.  The upper inset Figure shows the variation of the momentum scale $K$ of Eq.(\ref{K})
also as a function of $1/a_S k_F$.}
\end{figure}

%\textcolor{blue}
Unlike the case for lightcone fluctuations, where
$\Delta T\propto T$ \cite{ford} the saturation of $\Delta T$ here, and hence the vanishing of   $\Delta T/T$ for large $T$, makes comparison difficult. Nonetheless, the result $\Delta T = O(\epsilon_F^{-1})$ is as we would
expect by analogy  with quantum gravity \cite{hu}, as discussed in
\cite{ford,krein}. In quantum gravity, at best $(\Delta T/T)^2 \sim
\ell_P^2 \lambda_c^2 U  $ where $\ell_P$ is  the Planck length, and
$U$ is the  energy density of a bath of gravitons with a
characteristic wavelength $\lambda_c$. If, for example, we take the
energy density and typical wavelength of gravitons to be of the
order of those of microwave background radiation in the present
Universe, we find $\Delta T/T \approx 10^{-33}$, immeasurably small.
By analogy with gravity, on dimensional grounds $(\Delta T)^2$ can
be parameterized as $(\Delta T)^2 \sim U \omega_c / k_c^3 $  in
which the effective energy density $U$ is due to the condensate
fluctuations with a typical frequency $\omega_c$ and momentum $k_c$.
We estimate $U$ as $U \approx k_F^3 \epsilon_F$ with the frequency
$\omega_c \approx \bar{M}/\sqrt{\eta} \approx \epsilon_F$, and the
momentum $k_c \approx \sqrt{2 m \bar M^2/\rho_0} \approx k_F$ near
crossover regime, leading to the relatively large value of $\Delta T
=O(\epsilon_F^{-1})$ results.

In our model the speed of sound (\ref{c2}) vanishes in the BEC
regime because of the absence of direct diatomic self-interactions
in the Lagrangian density in (\ref{Lin}), but the qualitative
behaviour shown in Fig.1 does not rely on this fact. Suppose, as in
\cite{timmermans}, we include such a term
\be
L(\varphi) = - u_B |\varphi(x)|^4/4
\ee
\vskip -0.3cm
\noindent{in the integrand of (\ref{Lin}).} The
effect in $S_{eff}(\theta, \epsilon)$ of (\ref{SNONEQeff}) is just
to replace ${\bar M}^2$ by  ${\cal M}^2 = {\bar M}^2 + 6 u_B\kappa^2
|\varphi_0|^2$ in all results following (\ref{SNONEQeff}). [The term
linear in $\epsilon$, which corresponds to making the replacement
$\alpha G_0 \rightarrow \alpha G_0 + u_B\kappa |\varphi|^3$ in
(\ref{SNONEQeff}) has no effect, since it always contributes to
total derivatives in the calculations which follow.]
This leaves $c^2$ unchanged in the BCS regime because $\alpha\approx
0$ there but, since $\kappa^2 |\varphi_0|^2 \neq 0$  it permits
$c^2$ to tend to a non-zero limit in the deep BEC regime. However,
the vanishing of $\alpha$ in the deep BCS regime and the vanishing
of $\eta$ in the deep BEC regime are sufficient for fluctuations to
have no effect there. In the intermediate regime there will be a
reduction in $\Delta T$ due to the increase in ${\bar M}^2$ in Eq.(\ref{deltat/t6}) for the crossover regime. The effect is not dramatic but the details will depend on parameter choice.  Qualitatively the behaviour shown in Fig.1 will persist.

{\it Acknowledgements.---}This work was supported in part by the
Ministry of Science and Technology, Taiwan.

\end{document}